\documentclass[12pt]{article}
\usepackage{amssymb}
\usepackage{amsmath}
\usepackage{multirow}

\def\fun#1#2{\lower3.6pt\vbox{\baselineskip0pt\lineskip.9pt
\ialign{$\mathsurround=0pt#1\hfil##\hfil$\crcr#2\crcr\sim\crcr}}}
\newcommand{\beq}{\begin{equation}}
\newcommand{\eeq}{\end{equation}}

\usepackage{epsfig}
\DeclareMathOperator {\Res}{Res}
\newcommand{\be}{\begin{equation}}
\newcommand{\ee}{\end{equation}}
\newcommand{\bea}{\begin{eqnarray}}
\newcommand{\eea}{\end{eqnarray}}

\begin{document}
\title{Impact of the ionization of the atomic shell on the lifetime of the $^{229m}$Th isomer.}
\author{F. F. Karpeshin$^1$\footnote{e-mail: fkarpeshin@gmail.com}, M.B.Trzhaskovskaya$^2$ }

\maketitle
\begin{center}
$^1$D.I.Mendeleyev Institute for Metrology, St.-Petersburg, Russia  \\
$^2$National Research Center "Kurchatov Institute", Petersburg Nuclear Physics Institute, Gatchina  
\end{center}

    \begin{abstract}
      Recent experimental data are analyzed, concerning the half-lives of the $^{229m}$Th isomer in neutral atoms and various ions. 
Calculation is performed on the united platform of interplay of traditional and subthreshold resonance conversion. General agreement with experiment is obtained in the cases of Th I and Th III, a prediction is made concerning half-life in Th IV. 
Most critical is the case of Th II, where experimental data can be explained by interplay of various factors. A new physics is proposed,  based on dependence of the nuclear lifetime on the ambient conditions, such as atmospheric pressure. This must be taken into account in future experiments and their interpretation.   
\end{abstract}

\bigskip

\clearpage

\section{Introduction}

The idea of the combined atomic-nuclear transitions experienced  a period of strong development during the past decades. Making use of the resonance properties of the electron shell opens the way of manipulating the nuclei.  A possible practical realization of this way is the creation of the optical-nuclear clock based on a few-eVÿnuclear  isomer of $^{229}$Th \cite{pike,clok2}. For this purpose, more detailed information concerning the isomer properties is still needed,  including the exact value of the isomer energy and its half-life time.

      Indirect evidence of the presence of the isomeric level is known for decades. However, nobody could detect the isomer or its decay directly. Only recently, its decay through internal conversion (IC) was finally discovered \cite{lars}. Nevertheless, information about characteristic properties of the isomer, including its precise energy and lifetime, remains extremely scarce.   Thus, the estimate of its energy varies in time. An energy of 3.5 eV was considered for a long time \cite{reih}. Sometimes, a value of 5.5 eV was also used \cite{gimar}. Most recent measurements resulted in a higher value of 7.6$\pm$0.5 eV \cite{beck}. However, other values are also checked and cannot be excluded (e.g., \cite{lars,larsnew} and refs. cited therein). For the present purposes, we are oriented to this value as the latest data.

Moreover, the isomer half-life was measured for the first time. Its value of 7 $\mu$s was obtained in neutral atoms \cite{larsnew}, in coincidence with the theoretical estimate \cite{prc}. In neutral atoms, the isomer energy is  higher than the ionization potential $I_a$ = 6.3067$_3$ eV \cite{nist}. Therefore, decay occurs via IC. 
Let us start with consideration of this basic process ÿin more detail.

\section{Decay of the isomer through internal conversion in neutral atoms}
\label{ICna}

The decay width is described by 
\be
\Gamma=(1+\alpha(M1))\Gamma_\gamma^{(n)}    \,, \label{Gc}
\ee with $\alpha(M1)$ being the internal conversion coefficient (ICC), 
and $\Gamma_\gamma^{(n)}$ --- the radiative nuclear width.
Ground state electronic configuration is $(7s)^2(6d_{3/2})^2$. With the calculated ICC value in the 7s electronic shell,   $\alpha(M1) = 1.1\times 10^9$ \cite{prc}, Eq. (\ref{Gc}) allows one to conclude on the radiative nuclear half-life to be $T_{1/2} \approx$ 2 hours.  

It is didactic to trace how the estimation of the lifetime depends on the energy of the isomer.
Let us consider Fig. 2 in \cite{prc}. The ICC is exactly inversely proportional to $\omega^{-3}$, where $\omega$ is the transition energy. On the other hand, the nuclear radiative width $\Gamma_\gamma^{(n)}$ is proportional to $\omega^3$. Thus the expected lifetime, which is inversely proportional to  ICC times $\Gamma_\gamma^{(n)}$, holds with respect to variations of the supposed isomer energy. A change in the lifetime could be brought about by switching-on the next $6s$ shell to IC. But onset of the $6s$ shell would occur only at 37 eV in neutral atoms, and 56 eV --- in the singly charged ions. Currently such energies are not under discussion.

       In the ionized atoms of $^{229}$Th, IC becomes energetically closed. However, its mechanism remains effective in the form of bound internal conversion (BIC), also called resonance conversion because of its resonance character. Let us consider this process in more detail.

\section{Transition to the subthreshold region of BIC in the ions }

In the case of singly charged ions, the ionization potential is $I_a$ = 12.1 eV \cite{nist}, and the IC channel is energetically closed. However, deexcitation occurs mainly through many electronic bridges. For the first time this was shown in \cite{kabzon} for the 76-eV $^{235}$U isomer, and in  \cite{antib} in the case of neutral atoms of $^{229}$Th, under assumption of an isomer energy of 3.5 eV.  Calculations for singly charged ions of $^{229}$Th were performed in ref. \cite{PL2}.  More detailed calculations for the neutral atoms of $^{229}$Th, taking into account mixing of the electronic configurations, were undertaken in refs. \cite{prc,yaf}.  The results show that the main contribution comes from a few electronic transitions, in spite of the high fragmentation of the single-electron levels. For this reason, BIC remains to be a resonance process in its character, which does not exclude fluctuations of the lifetime depending on the accidental match of the energies of the nuclear and one of such a strong electron transition. With      a certain set of the basis electron configurations,  such a resonance enhancement was noted in \cite{prc,yaf}. As a result, it shortened the calculated lifetime by an order of magnitude.

      In Ref. \cite{atta}, the electronic bridges were considered for the 35-keV $M1$ transition in $^{125}$Te as the extension of traditional IC to the threshold case. The process received the name of BIC. Such an approach allows one to apply a conventional physical model and methods of calculations. The traditional ICC $\alpha(\tau,L)$ go over the analogues $\alpha_d(\tau,L)$, which are obtained by mere replacement of the conversion electron wavefunctions in the continuum by those from the discrete spectrum.  They acquire dimension of energy as a result of such a replacement. To obtain the dimensionless resonance conversion factor $R$, one has to multiply this analogue by the resonance  Breit---Wigner factor. Then we come to a conventional expression, similar to (\ref{Gc}):
\be
\Gamma=(1+R) \Gamma_\gamma^{(n)}	\label{lt}
\ee   
where, in turn,  the discrete conversion factor $R$ is expressed in terms of the analogue of ICC, $\alpha_d$:
\be
R = \sum_i\frac{\alpha_d^{(i)}(M1) \Gamma_t^{(i)} /2\pi} {(\Delta^{(i)})^2+(\Gamma_t^{(i)}/2)^2}  \label{Rg}
\ee 
with
\be
\Gamma_t^{(i)}=\Gamma_\gamma^{(n)}+\Gamma_a^{(i)}\approx \Gamma_a^{(i)}
\ee
being the total width of the nuclear and atomic decays, and
\be
\Delta^{(i)} = \omega_n-\omega_a^{(i)}
\ee --- the defect of the resonance. Summation in (\ref{Rg}) is over all the intermediate states. 

     Calculations were made within the framework of the multiconfiguration Dirac-Fock (MCDF) method, taking into account the interaction of the electronic configurations.      Concerning the electron configuration of the ground state in Th II, different information can be noted in the literature. One possibility is
      \be
   I: \qquad   (7s)^26d_{3/2} \quad J=3/2	\label{7s}  \ee 
(e.g., Ref. \cite{nist}), the other is
      \be
  II: \qquad    7s(6d_{3/2})^2 \quad J=3/2	\label{6d} \ee   (e.g., Ref. \cite{handbook}). Their  linear combination with comparable weights also may be used in practice. Our MCDF calculation shows that the ground state is (\ref{6d}). Nevertheless, these two single-electron states are practically degenerate. For this reason, there is a strong admixture of the (\ref{7s}) configuration in the ground state. In more detail,  the main components which comprise the wavefunction of the ground state are as follows: 
\begin{multline}
I: \qquad 0.709(  7s 6d_{3/2}^2)+0.157(7s^2 6d_{3/2})+0.037(6d_{3/2}^3)\\+
0.022  (8s\ 6d^2)+
0.020(6d_{3/2} 7p_{3/2} 5f_{5/2})+0.016(6d_{3/2}\ 6d_{5/2}^2)\\+
0.0082(7s 7p_{3/2} 5f_{5/2})+ 0.0077(7s 5f_{5/2}^2) + \ldots
\end{multline}
State (\ref{7s}) is comprised by the following main components:
\begin{multline}
II: \qquad 0.369(7s^2\ 6d_{3/2})+0.251(8s\ 6d_{3/2}^2)+
0.114(7s\ 6d_{5/2}^2)\\+0.086(7s\ 6d_{3/2}^2)+0.082(6d_{3/2}^3)+
0.024(6d_{3/2}\ 6d_{5/2}^2)\\+0.021(7s 7p_{1/2} 5f_{5/2})+
0.0076(7s 8s 6d_{3/2})+\ldots
\end{multline}

\section{Method of calculations and results for the ions}

	The consecutive calculation of influence of the electronic bridges on the lifetime was undertaken in ref. \cite{prc,yaf} for the neutral Th atoms, under assumption of the isomer energy of 3.5 eV. The calculations were performed within the MCDF method, taking into account electron configuration mixing. The resulting value of $R\approx 600$ was argued. Let us summarize the main experience of that investigation. 
At first, from a mathematical point of view, we note  the pole behavior of (\ref{Rg}). As a consequence,  the $R$ value depends on such parameters as the positions of the poles (more exactly, on the closeness to the nuclear energy) and their residues in the lower half-plane of the physical sheet.  The latters are expressed as follows:
\be
2\pi i \Res (R) \raisebox{-0.3ex} [0ex][-1ex]{ $\bigl|_{E=E_i}$} = \alpha_d^{(i)}\frac{\Gamma_i} {2\omega_i}\,.
\label{resid}
\ee
In the single-electron approximation, the value of $R$ is comprised by the electron transitions from the $7s$ to the  $8s$-, $9s$-, $10s$- and other $s$  states. They are decreasing inversely proportional to the ($\Delta^{(i)})^2$ values in the denominator of Eq. (\ref{Rg}). 
Accounting for a mixture of configurations leads to an increase in the number of poles and to their spread over the energy region, accompanied with a  decrease in the values of the residues. This is conventionally  known as fragmentation of the states. 

      In spite of the fragmentation, the number of the states which mainly contribute to the $R$ value remains limited.  They still lie in the vicinity of the single-electron  states $8s$, $9s$, $10s$ etc.  In neutral atoms, the resonance energy was approximately 3.5 eV due to the $7s-8s$ transition, which just coincided with the energy of the isomer of 3.5 eV assumed in \cite{prc,yaf}.
In the case of  Th II,  the $8s$ levels go up to 7 eV,  which value is essentially lower that the presently adopted nuclear energy. Next such strong states, descended from the $9s$ levels, are already at the energy of $\sim$8 -- 9 eV. 
However, there remains a number of levels in the vicinity of the nuclear transition energy, which are comprised by linear combinations of the configurations involved. Their contributions are   less strong.  As a result, the conversion amplitude becomes more regular with respect to the positions of the poles,  more smoothly depending on the transition energy. 

This is illustrated  in Table 1. For the deductive purposes, the $R$ factor is calculated for a \begin{table}[!tb]
\caption{\footnotesize Main intermediate states $i$ which make the predetermining contributions e total rate he isomer decay through the electronic bridges. For the purpose of illustration, the nuclear transition energy is taken as high as 8.99 eV, which value is very close to the calculated strong resonance at $\omega_i$ = 8.994 eV}
\begin{center}
\begin{tabular}{||c|c|c|c|c||}
\hline \hline
  $E_i$, eV  &     $J_i$  &   $\alpha_d^{(i)}$, eV  &   $\Gamma_\gamma^{(i)}$, eV  &    $R$ 	\\
  \hline   \hline
 6.053  &1.5   &   1.849$\times10^7$   &4.19$\times10^{-7}$  &   0.143	\\
   6.420 &  2.5    & 1.726$\times10^7$ &  5.60$\times10^{-7}$  & 0.233	\\
6.423  &    2.5    & 9.054$\times10^7$  & 3.35$\times10^{-7}$  & 0.732	\\
6.560 &  1.5     &  1302$\times10^7$ &  2.94$\times10^{-7}$  &  0.103	\\
   6.835& 1.5    &  1.351$\times10^8$  & 1.90$\times10^{-7}$  &  0.880	\\
6.867 &     2.5    & 2.644$\times10^7$   & 3.50$\times10^{-7}$ &  3.27	\\
6.891 & 2.5   &   5.999$\times10^8$ &  3.29$\times10^{-7}$   &    7.13	\\
7.213 &    2.5   &   2.221$\times10^7$ &  2.61$\times10^{-7}$ &  0.292	\\
7.508 &     1.5   &  1.022$\times10^8$  & 9.01$\times10^{-8}$ &  0.667	\\
7.523  &    2.5  &   2.462$\times10^7$ &  4.82$\times10^{-7}$  & 0.878	\\
   7.620 & 2.5   &   3.314$\times10^6$&   2.75$\times10^{-7}$ &  0.0773	\\
7.702 &  2.5    &   5.855$\times10^7$  &  2.66$\times10^{-7}$  & 1.49	\\
   7.849 &1.5   &   2.968$\times10^8$  &  3.88$\times10^{-7}$ &  14.1	\\
   7.995 &  1.5   &  8.802$\times10^8$  & 2.78$\times10^{-7}$ &  39.4	\\
8.032 &     2.5  &      1.288$\times10^7$  &   2.03$\times10^{-7}$  & 0.454	\\
8.994  &    2.5   &  3.142$\times10^8$   & 6.27$\times10^{-7}$  &  1.59$\times10^5$  	\\
9.019  & 1.5   &  1.106$\times10^8$   &   9.50$\times10^{-8}$  &   1950	\\
\hline \hline
\end{tabular}
\end{center}
\end{table}
representative isomer energy of 8.99 eV, which nearly coincides with one of the strongest states.  
Discrete ICC, widths and the resulting contributions are also presented for the other strongest states. Calculated value of $R$ in this case turns out to be $R=1.97\times10^5$. 
As one can see from Table 1, $\Gamma_i$ in Eq. (\ref{resid}) is a smooth function of the isomer energy $\omega_n$, weakly changing with the state. It is  $\alpha_d^{(i)}$  which varies by orders of magnitude from state to state. Therefore, we will mean the discrete ICC speaking about residues. Strong residues can be seen in  Table 1 at the energies of  6.891  and 7.995 eV. They correlate with the transitions to the 
$8s$-  and $9s$ single-electron states, respectively. The other of the listed residues are by 1 -- 2  orders of magnitude smaller. Moreover, there are still hundreds of levels unlisted, whose residues are yet smaller by 1 -- 3 orders of magnitude. 82 percent of the resulting $R$ value is due to one resonance state, and 98 percent are comprised by two of them, 7.995 and 7.849 eV . The other hundreds cannot compete because of  the longer distances from the resonance.

      The isomer energy of 7.6 eV just falls into this intermediate area.  The resulting $R$ value is $R$ = 1334. It is comprised by combination of  hundreds of the same electronic configurations. It is, however, only five  of them, whose approximately equal contributions exhaust 75 percent of the resulting $R$ value, as presented in Table 2.  Their main structure is also listed in terms of nonrelativistic basis configurations.  Analysis of the listed values demonstrates softening of the pole behavior of the $R$ factor by influence of  the electronic structure of the levels. Indeed, the levels  have very different defects of resonance $\Delta^{(i)}$, which vary from 0.02 eV for the second level to 0.4 eV for the fifth level.   This means that their contributions which are 
$\sim \Delta^{-2}$, according to (\ref{Rg}), might vary by 2 -- 3 orders of magnitude. But they remain comparable, which reflects the importance of the electronic structure.

\section{Comparison to experiment}

Baring in mind the calculated value of the BIC factor $R$ = 1334, the half-life time calculated by means of (\ref{lt}) turns out to be about 6  s. 
 \begin{table}[!tb]
\caption{\footnotesize Main configurations contributing to the electronic bridges, and the related $R$ values, for the nuclear transition energy $\omega_n$=7.6 eV. The total contribution of the listed configurations comprise 70\% of the total value of $R$ = 1334}
\begin{center}
\begin{tabular}{||c|c|c|c|c|c|c||}
\hline \hline
$E_i$, eV  &    $J_i$    &   $|\Delta|$, eV &  $\alpha_d^{(i)}$, eV  &   $\Gamma_\gamma^{(i)}$, eV   &    composition   &	$R$  \\
\hline
7.523 &   2.5   &   0.08& 4.076$\times10^7$  & 2.05$\times10^{-7}$ & 0.64\ $(6d7p5f)+0.11\ (7s5f^2)+0.10\ (8s6d^2)$  &  226	\\
7.620 &2.5  &    0.02 &  5.486$\times10^6$  & 1.06$\times10^{-7}$  & $0.36\ (6d7p5f)$ + 0.14\ $(7s8s6d)
+0.13\ (8s6d^2)$   & 238	\\
7.702  & 2.5  &  0.1 & 9.691$\times10^7$  & 1.11$\times10^{-7}$ & $0.41\ (7s5f^2)+0.34\ (6d7p5f)$ & 164		\\
7.849 &    1.5   &  0.25 & 4.912$\times10^8$ &  1.59$\times10^{-7}$  &0.50\ (6d7p5f)$ + $0.28\ $(7s8s6d)$ & 202	\\
7.995 & 1.5    &  0.4 & 1.457$\times10^9$ &  1.11$\times10^{-7}$  & 0.61\ $(7s8s6d)$+0.26\ $(6d7p5f)$   & 164	\\
\hline \hline
\end{tabular}
\end{center}
\end{table}
Surprisingly, the calculated half-life looks to be  in contradiction with experimental data \cite{larsnew}. The latter show a value which is 10 ms or shorter. That is at least by 2 -- 3 orders of magnitude less than the theoretical estimate. Let us  search for probable reasons for such contradiction. 

1. Qualitatively, were configuration (\ref{7s}) assumed, a shorter half-life by a factor of about 2 could be expected. Actually, our calculation under assumption of the ground electronic configuration (\ref{7s}) yields in the  value of $R$ = 6933.   From these estimations, one can again conclude on a strong dependence of the Th {II} lifetime on the electron configuration.  The latter value is in better agreement with  experiment.  However, this  results in the lifetime of about 1 s, which is still too much in comparison with experiment. 

2. Yet unexplained difference may be interpreted as a consequence of a randomly close coincidence with a strong electron level. Such an example was considered in ref. \cite{prc}, when a random coincidence within 0.03 eV led to the change of the $R$ factor by an order of magnitude. 
\begin{table}[!tb]
\caption{\footnotesize Numerical simulations of the isomer lifetime in cases of various representative nuclear energies $\omega_n$ close to 7.6 eV. The energies are chosen as being in resonance with various electronic transitions}
\begin{center}
\begin{tabular}{||c|c|c|c|c|c||}
\hline \hline
&\multicolumn{4}{c|}{ Main intermediate state } & \\
\cline{2-5}	
$\omega_n$, eV& $E_{lev}$, eV & $J$ & Leading configuration &  $R_j$	 & Lifetime, s  \\
\hline  \hline
6.83&6.835&1.5&0.84$(8s6d^2)+0.0 3(7s8s6d)$ & 72655	 & 0.10 \\
7.25&7.248&1.5& 0.65$(8s6d^2)+0.16(7s8s6d)$ & 14213	 & 0.51  \\
7.27&7.302&1.5&0.80$(5f6d7p)+0.04(7s8s6d)$ &1022  &  7.0	\\
7.32&7.322&2.5&0.30$(8s6d^2)+0.21(7s8s6d)+0.26(5f6d7p)$ &32674  & 0.22	\\
7.51&7.508&1.5&0.43$(8s6d^2)+0.30(7s8s6d)$ & 194980	&  0.037	\\
7.56&7.561&2.5&0.18$(8s6d^2)+0.39(7s8s6d)$ & 100130  &  0.072	\\
7.86&7.849&1.5&0.29$(7s8s6d)+0.43(5f6d7p)$&2289  &  3.1  \\
8.01 & 7.995 &1.5&0.61$(7s8s6d)+0.23(5f6d7p)$ & 720000 & 0.01 \\
8.28  & 8.280 & 1.5 & 0.66$(8s6d^2)+0.03(7s8s6d)+ 0.14(5f6d7p)$ & 4905200 & 0.0015 \\
\hline \hline
\end{tabular}
\end{center}
\end{table}
The most important  levels are those at 6.891,  7.995 and 8.994 eV. The results, calculated for  several representative isomer energies close to the resonances, are shown in Table 3. In the case of $\omega_n$ = 8.28 eV, the main contribution comes from the  level at 7.995 eV.
In this case, the resonance condition looses, and agreement with experiment for the calculated half-life of 0.01 s is achieved with $\Delta$ = 0.015 eV. 

	Quantitatively, the probability that  $R$ factor will not be less than $R_c$ can be defined as follows:
\be
w(R)\raisebox{-0.3em}[0em][-1em] {$\bigl|_{R>R_c}$}=\frac1{\Delta \omega_n}\int\limits_{(\Delta \omega_n)} H\left(R(\omega)-R_c\right)  \,,	\label{Rc}
\ee
where $ R(\omega)$ is the $R$ factor for the nuclear isomer energy $\omega$, $H(x)$ --- the Heaviside step function, defined as follows:
\bea
H(x) \, = \,   \left\{
            \begin{array}{l@{\hspace{2cm}}l}
                 1  &   \mbox{for } x \geq 0  \\
                 0    &  \mbox{for } x < 0
            \end{array}
   \right.
\eea
	Integration in Eq. (\ref{Rc}) is performed over the assumed domain of the nuclear energy. Inserting $R_c=7.2\times10^5$ in Eq. (\ref{Rc}),  values of $w(R)\raisebox{-0.3em}[0em][1em]{$\bigl|_{R>R_c}$}$ = 0.10 and 0.32 have been obtained, using either (\ref{6d})  or (\ref{7s}) the ground state function, respectively. In more detail, one third of the total value is due to the contribution from the 7.99 eV state. The received value is not small. One can say that it reconciles the theory with the experiment.

      3. Searching for  further factors, we note that Eq. (\ref{Rg}) deserves a remarkable comment. It is conventionally accepted that the nuclear properties, specifically the radioactive decay constant, are essentially independent of the physical environment. At most, variations of the nuclear lifetimes are mentioned depending on the chemical environment. They were found not to exceed the level of  ten percent. The variations may arise due to change of the IC rate with the variation of the population of the upper electronic shells. Such a stability against the environmental medium  underlies the idea of the nuclear clock. 
In a certain sense,  somewhat aside is a number of papers, where it was shown that the nuclear processes may be strongly affected by means of laser radiation, using electron shells as resonators \cite{kabzon,PL1,canad,Chin,Hf09,echa,book}. Specifically, it was predicted  that the decay of the $^{229}$Th isomer  would be by $\sim$700 times faster in the hydrogen-like ions \cite{zylic}. A similar effect was also predicted in $^{169}$Yb \cite{YbZETP}. But in fact, Eq. (\ref{Rg}) says much more. Namely, as one can see from Eqs. (\ref{lt}) -- (\ref{Rg}), in the case of BIC the atomic width $\Gamma_a^{(i)}$ enters the expression for the nuclear decay width directly as a factor, on the equal footing with the radiative nuclear width and the discrete analogue of  ICC $\alpha_d^{(i)}$. That is, the nuclear decay width turns out to be directly proportional to the atomic width $\Gamma_a^{(i)}$, which in turn is determined by completely different factors, including such as mere temperature and pressure of the physical environment. 
As is known, at normal conditions, due to the collisional broadening, the full atomic line widths can be by two orders of magnitude as large as the natural line width \cite{broad1,broad2}.  
Turning to the concrete conditions of experiment \cite{lars,larsnew}, we know that 10 ms is just the time during which fresh atoms and ions of isomeric $^{229m}$Th are kept in the stopping cell for the purpose of thermalization. The cell is filled by helium buffer gas at the pressure of 40 mbar. This comprises 1/10 of a normal atmosphere pressure. Therefore, taking into account what is said above, the lifetime of the excited atomic state might be reduced by up to an order of magnitude. Allowance for this circumstance may diminish our theoretical estimate to 0.6 and 0.1 s in the cases of configurations (\ref{6d}) and (\ref{7s}), respectively. Therefore, a value of 
$R =R_c=7.2\times10^4$ will be enough for explanation of the experiment. This is achieved with the probabilities of   $w(R)\raisebox{-0.3em}[0em][1em]{$\bigl|_{R>7.2\times10^4}$}$ = 32 and 41 percent, for (\ref{6d}) and (\ref{7s}) ground state configurations, respectively. These values may satisfy any  comparison between theory and experiment at the contemporary stage of investigation. 
We summarize the results in Table 4.
\begin{table}[!tb]
\caption{\footnotesize Calculated probabilities that a value of $R$ is $R>R_{c}$ as a result of a close coincidence of the nuclear and electronic transition energies under assumption that the nuclear isomer energy $\omega_n$  may be uniformly distributed over   the accepted domain  of 7.6$\pm$0.5 eV. The listed values of $R_c$ reply to various possibities of how the nuclear half-life less than 0.01 s can be realized (see text)}
\begin{center}
\renewcommand{\multirowsetup}{\centering}
\newlength{\LL} \settowidth{\LL}{$R_c$}
\begin{tabular}{||c||c|c||}
\hline  \hline
\multirow{2}{\LL}{$R_c$} & \multicolumn{2}{c||}{$w(R)\raisebox{-0.3em}[0em][1em]{$\bigl|_{R>R_c}$}$\,, \%}   \\
\cline{2-3}
   & $7s(6d_{3/2})^2$ &   $(7s)^26d_{3/2}$   \\
\hline  \hline
$7.2\times10^5$  &   10     &   13    \\
$7.2\times10^4$  &  32   &  41  \\
 \hline  \hline
\end{tabular}
\end{center}
\end{table}

      In the case of Th {III}, the ground state configuration is $6d_{3/2}5f_{5/2}$ \cite{nist,austral}.  Therefore, the electronic bridge through the transition $7s-8s$ is ruled out in the main approximation. According to what is said above, it may occur due to admixture of the $7s$, $8s$ and other $s$ states to the main configuration in the ground state.  Therefore,  the mixing coefficients comprise amplitudes of tenths at most, and the related $R$ values are suppressed by a factor of 10 at least. Our calculation shows that this
reduces the expected  $R$ value in comparison with the previous case of the single ions, to the values of   $R^{III}$ = 2.8. This results in  the expected half-life of about 40 minutes. It thus becomes a step closer to the radiative half-life. This is in accordance with the observed value \cite{lars} of $T_{1/2} >$ 100 s. The latter value allowed to conclude that the energy of the isomer is less than $\omega_n \lesssim I_a$ = 18.3 eV for these ions \cite{lars}.

      In the three-fold ions of Th {IV}, the calculated ground state is $5f_{5/2}$. The $R$ factor further diminishes. Its value obtained is $R$ = 0.08. Therefore, the expected lifetime approaches  that in the bare nuclei, that is about 2 hours.

      \section{Conclusion}

We performed systematic analysis of the calculated lifetimes of the $^{229m}$Th isomer in ions of different degree of ionization, from neutral atoms to three-fold ions. The consideration is conducted on the unified platform of  interplay of traditional IC with resonance BIC,  embodied by electronic bridges. Such an approach points out clearly that the lifetime is not so critical to exact values of the isomer energy, when the latter is within the domains above or lower than the potential of ionization $I_a$ of the atoms. It is mainly defined by the interrelation between the isomer energy and the ionization potential. In the neutral atoms, the decay rate is determined by internal conversion. Contrary, the nuclear transition energy is below the threshold of internal conversion in the ionized atoms. In these atoms, the deexcitation occurs through the subthreshold IC, or BIC. The important  factor in these cases is a contribution from the $7s-8s$ or $7s-9s$ single-electron transition. Such a transition may regularly occur in the ions of Th II, where the ground level is either (\ref{7s}), or (\ref{6d}). The role of the electronic bridge is expected to be much lower in the Th III and Th IV atoms, where the ground state configuration  is $6d5f$ and $5f$, respectively, and  the $7s$ single-electron state is absent in the main terms. As a consequence, the half-life in these ions is expected to increase steeply, approaching that in bare nuclei.

      	The results obtained are compared to available experimental data for the neutral atoms, one- and two-fold ionized atoms. 
First, worthy of noting is the agreement which is achieved in the case of neutral atoms. Analysis of this coincidence is made in section \ref{ICna}. Properly speaking, the lifetime is not mastered by the radiative nuclear lifetime as related to the isomer energy. But it is determined  by the Weisskopf hindrance factor.  With the same hindrance factor, one can change the isomer energy by several times, but the lifetime essentially holds. Measured lifetime coincides with the theoretical one, based on the typical hindrance factor for the nuclei in this area of $\sim$300  and the calculated ICC \cite{prc}, which is 1.1$\times10^9$ at the isomeric energy of 7.6 eV \cite{larsnew}. An estimate of approximately two hours follows for  the nuclear radiative lifetime.  

      Furthermore, assuming this energy value, the calculated lifetime in the two-fold ionized atoms also turns out to be in agreement with the experiment. In this case, the ground-state configuration contains no more the $7s$ single-electron level in the leading term. As a result, the $R$ factor of BIC is $R \approx$ 3, and the lifetime approaches 40 minutes.  This value does not contradict the experiment, which yet gives only a lower bound of $\sim$ 2 min \cite{lars, larsnew}.

      More critical is the case of Th II, where comparison of theory with experiment is also possible.  We advanced the results \cite{prc,yaf} concerning fragmentation of the electronic levels brought about by the configuration interaction.  We separated out from one another the effects, arising due to the atomic structure (residues), and those caused by the pole-like resonance enhancement.  Both are important for correct calculation of the final $R$ value. The present consideration  confirms that in spite of the tremendous number of states and strong their mixing, generally it is only a few of them which mainly determine the RC rate.     
Such states concentrate around the energies of 6 --7 and 8 -- 9 eV, descending from the $7s-8s$ or $7s-9s$ single-electron transitions, respectively.  The assumed nuclear isomeric energy of 7.6 eV falls just between these values. In this region, the pole-like behavior is smeared. This results in the expected value of the BIC factor $R$ = 1334, and the related lifetime of about 1 s in the case of (\ref{7s}) ground-state configuration, which better agrees with the experiment. For the case of (\ref{6d}), it increases up to 6 s. At the same time, the experiment shows a value by a factor of  at least 100 times shorter.  Such a deviation, after a fair agreement in the  two previous cases, looks all the more intriguing. 	Discussing  ways of improving agreement with experiment, it is noted that, first, the decay rate in Th II atoms may be raised essentially in the case of coincidence of the nuclear transition energy with the electronic one. Supposing the uniform distribution of a probable  nuclear energy within the interval of $7.6\pm 0.5$ eV, we estimated the probability of such a coincidence, listed in Table 4. The probability  is quite high in the case of the ground-state configuration (\ref{7s}): the confidence interval (the total $\Delta$ value, satisfying condition (\ref{Rc})) covers up to 23 percent of the  uncertainty of the nuclear energy. 

   	Another circumstance, which enlarges the probability, is related with new physics. It is suggested that a moderate collisional 
broadening by up to an order of magnitude can be expected due to the experimental conditions \cite{lars,larsnew}, where fresh ions are kept in the stopping cell at the pressure of 40 mbar. This broadening may shorten the lifetime of the atomic levels populated by the BIC electrons by an order of magnitude. In turn, this also shortens the isomeric lifetime by an order of magnitude. Account of this mechanism allows one to resolve the question of  agreement between theory and experiment at a higher  level of confidence of  up to 74 percent. This mechanism of nuclear sensitivity to the ambient conditions is not surprising, in view of the fact that BIC from the very beginning was designed for the purpose of inducing nuclear isomer decay \cite{kabzon}. It deserves the closest attention in the future experiments. 

    Furthermore, a prediction is made concerning the half-life in the three-fold ions. Due to a very simple $5f$ configuration of the ground state, the $R$ factor becomes less than unity, and the lifetime approaches that in bare nuclei. We note a beautiful analogy between our model of variation of the half-lives with ionization, as obtained above, with that for 45 to 48 fold ions of  $^{125}$Te \cite{atta}. In both cases, the lifetime regularly changes as usual internal conversion gives drive to subthreshold resonance conversion, which in turn fades away with further ionization.

      Summarizing, there is no contradiction at the time being with experiments \cite{lars,larsnew}. Future research into the singly ionized atoms should be directed to, first, getting more precise data on the nuclear isomer  energy, and also electronic spectroscopy around the $8s$ and $9s$ levels. Second, a note should be taken of influence of the ambient conditions on the nuclear half-life in the case of BIC.
 
\bigskip
The authors would like to acknowledge inducing highly heuristic discussions of his experimental results with L. Wense. They are also grateful to L. F. Vitushkin for helpful discussions, M. Okhapkin and  A. Popov for fruitful remarks.

\end{document}